\documentclass[twocolumn,prd,superscriptaddress,altaffilletter,nofootinbib,prep,preprintnumbers,longbibliography]{revtex4-1}
\usepackage{amsmath}
\usepackage{amssymb}
\usepackage{amsfonts}
\usepackage{comment}
\usepackage{graphicx}
\usepackage[
    colorlinks=true,
    linkcolor=BrickRed,
    citecolor=NavyBlue,
    urlcolor=NavyBlue,
    breaklinks=true
]{hyperref}
\usepackage{acronym}
\usepackage[usenames,dvipsnames]{xcolor}
\usepackage{xspace}
\usepackage{tikz}
\usepackage{siunitx}
\usepackage{makecell}
\usepackage{multirow}
\usepackage{bm}
\usepackage{romannum}
\usepackage{orcidlink}
\usepackage[caption=false]{subfig}
\usepackage{booktabs}
\usepackage{physics}
\usepackage{tensor}
\usepackage{braket}
\usepackage{placeins}
\usepackage[utf8]{inputenc}

\usetikzlibrary{arrows,shapes,trees,decorations.pathreplacing}
\allowdisplaybreaks

\newcommand{\gw}[0]{\ac{GW}\xspace}
\newcommand{\gws}[0]{\acp{GW}\xspace}
\newcommand{\lvk}[0]{\ac{LVK}\xspace}
\newcommand{\gr}[0]{\ac{GR}\xspace}
\newcommand{\bbh}[0]{\ac{BBH}\xspace}

\newcommand{\snr}[0]{\ac{SNR}\xspace}
\newcommand{\bh}[0]{\ac{BH}\xspace}
\newcommand{\qnm}[0]{\ac{QNM}\xspace}
\newcommand{\qnms}[0]{\acp{QNM}\xspace}
\newcommand{\imr}[0]{\ac{IMR}\xspace}
\newcommand{\psd}[0]{\ac{PSD}\xspace}
\newcommand{\kde}[0]{\ac{KDE}\xspace}
\newcommand{\hpd}[0]{\ac{HPD}\xspace}
\newcommand{\ovl}[0]{\ac{OVL}\xspace}
\newcommand{\ovls}[0]{\acp{OVL}\xspace}

\begin{document}
\pagenumbering{arabic}

\title{Ringdown analysis of GW250114 with orthonormal modes}

\author{Motoki Suzuki \orcidlink{0009-0009-3585-0762}}
\affiliation{Department of Physics, Graduate School of Science, The University of Tokyo,
7-3-1 Hongo, Bunkyo-ku, Tokyo 113-8655, Japan}
\affiliation{Institute for Cosmic Ray Research, The University of Tokyo, 
5-1-5 Kashiwanoha, Kashiwa, Chiba 277-8582, Japan}
\author{Kei-ichiro Kubota \orcidlink{0000-0002-1576-4332}}
\affiliation{Institute for Cosmic Ray Research, The University of Tokyo, 
5-1-5 Kashiwanoha, Kashiwa, Chiba 277-8582, Japan}
\author{Soichiro Morisaki \orcidlink{0000-0002-8445-6747}}
\affiliation{Institute for Cosmic Ray Research, The University of Tokyo, 
5-1-5 Kashiwanoha, Kashiwa, Chiba 277-8582, Japan}
\author{Hayato Motohashi \orcidlink{0000-0002-4330-7024}}
\affiliation{Department of Physics, Tokyo Metropolitan University, 1-1 Minami-Osawa, Hachioji, Tokyo 192-0397, Japan}
\author{Daiki Watarai
\orcidlink{0009-0002-7569-5823}}
\affiliation{Research Center for the Early Universe (RESCEU), Graduate School of Science, The University of Tokyo, Tokyo 113-0033, Japan}

\begin{abstract}

GW250114 is the loudest gravitational wave event to date observed by LIGO-Virgo-KAGRA Collaboration.
Owing to its high \snr, previous analyses based on \qnm superpositions have suggested evidence of the fundamental and the first overtone of the $\ell=m=2$ mode in this event.
However, \qnms are not orthogonal and the inclusion of multiple \qnms induces correlations among them, which can hinder the robust identification of subdominant \qnms.
To address this challenge, we apply an analysis based on orthonormalized \qnms~[\href{https://journals.aps.org/prd/abstract/10.1103/lhkp-w6hm}{S. Morisaki \textit{et al.}, Phys.\ Rev.\ D \textbf{112}, 124083 (2025)}] to GW250114.
We find that, in the model including three $\ell=m=2$ \qnms up to the second overtone, the first overtone of the $\ell=m=2$ mode is more strongly supported than in previous nonorthogonal analyses, with the inferred significance increasing from 86.2\% to 99.9\%.
Furthermore, we estimate deviations from the Kerr prediction using the orthonormal \qnm framework and find no significant deviation, consistent with previous analyses.
These results demonstrate that the orthonormal \qnm framework provides a more robust way to identify subdominant modes in high-\snr ringdown signals, highlighting its potential for future gravitational wave observations.

\end{abstract}

\maketitle

\acrodef{GW}{gravitational wave}
\acrodef{LVK}{LIGO-Virgo-KAGRA}
\acrodef{GR}{general relativity}
\acrodef{BBH}{binary black hole}
\acrodef{SNR}{signal-to-noise ratio}

\acrodef{QNM}{quasinormal mode}
\acrodef{BH}{black hole}
\acrodef{IMR}{inspiral-merger-ringdown}
\acrodef{PSD}{power spectral density}
\acrodef{KDE}{kernel density estimator}
\acrodef{HPD}{highest posterior density}
\acrodef{OVL}{overlap coefficient}

\acresetall 

\section{Introduction}

The first direct detection of \gws~\cite{LIGOScientific:2016aoc} marked the beginning of \gw astronomy.
Following this milestone, \lvk Collaboration has detected more than 300 \gw events up to its fourth observing run~\cite{LIGOScientific:2018mvr,LIGOScientific:2020ibl,LIGOScientific:2021usb,KAGRA:2021vkt,LIGOScientific:2025slb}.
Most of these signals are consistent with \gws emitted by \bbh mergers.
After the merger, a highly distorted \bh is formed, which settles into a single rotating \bh while emitting \gws.
This phase is known as the ringdown, and its signal can be modeled in \gr as a superposition of damped sinusoids called \qnms.
Each \qnm is labeled by three indices $(\ell,m,n)$, where $\ell$ and $m$ are angular indices and $n$ denotes the overtone number.
The frequency and damping time of each \qnm are uniquely determined by the mass and spin of the remnant Kerr \bh, as a consequence of the no-hair theorem.
Therefore, the ringdown signal encodes rich information about the remnant \bh and the underlying theory of gravity~
\cite{LIGOScientific:2019fpa, LIGOScientific:2020tif, LIGOScientific:2021sio, LIGOScientific:2026qni, LIGOScientific:2026fcf, LIGOScientific:2026wpt, LIGOScientific:2016lio, Isi:2019aib, LIGOScientific:2025rid, LIGOScientific:2025wao}.

Owing to the no-hair theorem, the simultaneous detection of multiple \qnms in a single \gw event enables independent measurements of the mass and spin of the remnant \bh from each mode, as well as a consistency test among them.
This approach, referred to as \bh spectroscopy, provides a relatively model-independent test of \gr~\cite{Dreyer:2003bv, Berti:2005ys, Berti:2007zu, Berti:2025hly}.

Several studies have reported possible evidence for subdominant \qnms, in addition to the dominant $(2,2,0)$ mode, in some \gw events.
A notable example is GW150914~\cite{LIGOScientific:2016aoc}, the first detected \gw event.
Reference~\cite{Isi:2019aib} reported evidence for the presence of at least one overtone ($n>0$) of the $(\ell,m)=(2,2)$ mode with $3.6\sigma$ confidence, obtained by starting the analysis at the peak of the signal.
However, this analysis is based on modeling the signal as a superposition of linear \qnms.
Near the peak of the signal, additional contributions such as higher overtones, nonlinear \qnms~\cite{Ma:2024qcv, London:2014cma, Cheung:2022rbm, Khera:2023oyf, Mitman:2022qdl, Khera:2024bjs}, and direct waves~\cite{Oshita:2025qmn, Lu:2025vol} may be present, and the validity of this approximation is, therefore, not well established.
Several studies have revisited this event, reporting in some cases lower significance for the presence of overtones~\cite{Cotesta:2022pci, Ma:2023cwe, Correia:2023bfn, Wang:2024yhb, Wang:2021elt, Isi:2022mhy, Finch:2022ynt, CalderonBustillo:2020rmh, LIGOScientific:2020tif, Chandra:2025ipu, Lu:2025mwp}.
In parallel, the validity of fits based on a superposition of \qnms near the signal peak has been investigated in a number of studies~\cite{Baibhav:2023clw, Giesler:2019uxc, Cheung:2023vki, Forteza:2021wfq, Dhani:2020nik, Takahashi:2023tkb, Finch:2021iip, Giesler:2024hcr, Mitman:2025hgy, Bhagwat:2019dtm, Gao:2025zvl, Nee:2023osy, Clarke:2024lwi}.
In light of these studies, since the onset of the linear perturbative regime remains unclear, the presence of subdominant modes is often discussed at times sufficiently after the peak~\cite{LIGOScientific:2025rid, LIGOScientific:2025wao}.

Evidence for subdominant modes with different angular indices has also been reported.
GW190521~\cite{LIGOScientific:2020iuh, LIGOScientific:2020ufj} is a high-mass \bbh merger with an asymmetric mass ratio, for which Ref.~\cite{Capano:2021etf} has reported evidence for the $(3,3,0)$ mode, while Ref.~\cite{Siegel:2023lxl} has reported evidence for the $(2,1,0)$ mode.
Another example is GW231123~\cite{LIGOScientific:2025rsn}, the most massive \bbh event to date.
Its inferred mass in the pair-instability mass gap, combined with its possibly high spin, has attracted significant interest in its formation scenario~\cite{Bartos:2025pkv, Paiella:2025qld, Kiroglu:2025vqy, Croon:2025gol, Li:2025fnf, Passenger:2025acb, Liu:2025ogx, Stegmann:2025cja, Li:2025pyo}.
Several studies have suggested the presence of \qnms beyond the dominant $(2,2,0)$ mode~\cite{LIGOScientific:2025rsn, Siegel:2025xgb, Wang:2025rvn}.
However, alternative interpretations beyond the standard stellar-origin \bbh scenario~\cite{Yuan:2025avq, Cuceu:2025fzi, Shan:2025dcd}, as well as uncertainties in the inference of source properties due to waveform systematics and detector noise~\cite{Ray:2025rtt, Hu:2025lhv, Bini:2026kwz}, have been discussed, making the interpretation of these \qnms less straightforward.

In principle, the linear perturbative regime can be described as a superposition of an infinite number of linear \qnms.
In practice, however, analyses are performed using template waveforms that include only a finite number of \qnms contributing dominantly to the signal.
Since \qnms are not orthogonal, correlations arise among these amplitudes, obscuring the identification of individual modes in the data~\cite{CalderonBustillo:2020rmh, Chandra:2025ipu, Clarke:2024lwi}.
Furthermore, as more modes are included, these degeneracies become more pronounced, leading to increased computational cost.
These challenges are expected to become more significant with the advent of next-generation detectors, such as the Cosmic Explorer~\cite{Reitze:2019iox} and the Einstein Telescope~\cite{Punturo:2010zz}, whose improved sensitivity will allow the detection of additional subdominant modes.

To address these challenges, we adopt the semianalytic method based on orthonormalized \qnms, which reduces parameter degeneracies and accelerates computations~\cite{Morisaki:2025gyu}.
In this paper, we apply this approach to the event GW250114\_082203, henceforth GW250114~\cite{LIGOScientific:2025rid}.
This event was detected by the LIGO detectors~\cite{LIGOScientific:2014pky} on January 14, 2025, and has the largest network \snr to date, with a value of approximately 80~\cite{LIGOScientific:2025rid, LIGOScientific:2025wao}.
At the time of this event, the Virgo~\cite{VIRGO:2014yos} and KAGRA~\cite{10.1093/ptep/ptaa125} detectors were not in operation.
Reference~\cite{LIGOScientific:2025rid} reports an \imr analysis using the \textsc{NRSur7dq4} waveform model~\cite{Varma:2019csw}, consistent with a \bbh merger with source-frame component masses of $m_1=33.6^{+1.2}_{-0.8}\,M_{\odot}$ and $m_2=32.2^{+0.8}_{-1.3}\,M_{\odot}$ and dimensionless spin magnitudes of $\chi_1\leq0.24$ and $\chi_2\leq0.26$.
The source-frame mass and spin of the remnant \bh are estimated to be $M^{\mathrm{src}}_f=62.7^{+1.0}_{-1.1}\,M_{\odot}$ and $\chi_f=0.68^{+0.01}_{-0.01}$, respectively.
In the ringdown of GW250114, previous studies~\cite{LIGOScientific:2025rid, LIGOScientific:2025wao} have reported evidence of the subdominant $(2,2,1)$ mode, in addition to the dominant $(2,2,0)$ mode, at analysis start times of several $t_{M_f}$ after the signal peak, where $t_{M_f}$ corresponds to a characteristic timescale of $0.337\,\mathrm{ms}$ for this event.
However, in that time range, including the $(2,2,2)$ mode in addition to these modes in a template waveform, i.e., extending the 220+221 model to the 220+221+222 model, leads to discrepancies in the estimated amplitudes of individual \qnms between the two models.

In this paper, we perform semianalytic analyses based on orthonormalized \qnms~\cite{Morisaki:2025gyu}.
For comparison, we also perform analyses based on nonorthogonal \qnms using the \textsc{ringdown} package~\cite{Isi:2021iql}, following previous studies~\cite{LIGOScientific:2025rid, LIGOScientific:2025wao}.
In the semianalytic analysis, the posterior distributions of the amplitudes remain largely unchanged under the inclusion of additional modes.
In particular, the posterior distribution of the $(2,2,1)$ mode shows little change when extending the 220+221 model to the 220+221+222 model.
As a result, at $6\,t_{M_f}$ after the peak, the posterior of the $(2,2,1)$ mode amplitude excludes zero at the 86.2\% level in the nonorthogonal \qnms analysis, whereas it excludes zero at the 99.9\% level in the semianalytic analysis for the 220+221+222 model.
In addition, using orthonormalized \qnms, we estimate deviations of the complex frequency from Kerr prediction for the $(2,2,1)$ mode.
We find that the inferred deviations are consistent with those obtained using the \textsc{ringdown} package.

\section{Method}

\subsection{Signal model and inference method}

The ringdown strain can be written as
\begin{align}
    h(t)&=h_+(t)-ih_{\times}(t) \notag\\
    &=\sum_{\ell mn}\qty(
        \mathcal{A}_{\ell mn}e^{-i\tilde{\omega}_{\ell mn}t}
        +\mathcal{A}^{\prime}_{\ell mn}e^{i\tilde{\omega}^{\ast}_{\ell mn}t}
    ),
    \label{eq:complex_strain}
\end{align}
where $\ell\geq2$, $-\ell\leq m\leq\ell$, and $n\geq0$ are the multipole, azimuthal, and overtone numbers, respectively.
$\mathcal{A}_{\ell mn}$ and $\mathcal{A}^{\prime}_{\ell mn}$ are complex amplitudes.
The complex frequency $\tilde{\omega}_{\ell mn}\equiv\omega_{\ell mn}-i/\tau_{\ell mn}$ depends on the remnant \bh's mass $M_f$ and spin $\chi_f$.
Note that in this summation, the terms with $m>0$ correspond to prograde modes, while the terms with $m<0$ correspond to retrograde modes.
Using Eq.~\eqref{eq:complex_strain}, the strain measured by the $I$th detector is
\begin{align}
    h^I(t)&=F^I_+h_++F^I_{\times}h_{\times}=\sum_{\ell mn}\sum_{j=0}^3 c_{j,\ell mn}v^I_{j,\ell mn}(t),
\end{align}
where $F^I_+$ and $F^I_{\times}$ are the antenna pattern functions of the $I$th detector.
The coefficients $c_{j,\ell mn}$ are defined as
\footnote{
    The definition of the coefficients differs from that in Ref.~\cite{Morisaki:2025gyu}, owing to a different convention for the complex strain, $h=h_+\pm ih_{\times}$.
    In this work, we adopt the convention in Eq.~\eqref{eq:complex_strain}, consistent with the \textsc{ringdown} package.
}
\begin{subequations}
    \begin{align}
        c_{0,\ell mn}&=\Re(\mathcal{A}_{\ell mn}+\mathcal{A}^{\prime}_{\ell mn}), \\
        c_{1,\ell mn}&=\Im(\mathcal{A}_{\ell mn}-\mathcal{A}^{\prime}_{\ell mn}), \\
        c_{2,\ell mn}&=-\Im(\mathcal{A}_{\ell mn}+\mathcal{A}^{\prime}_{\ell mn}), \\
        c_{3,\ell mn}&=\Re(\mathcal{A}_{\ell mn}-\mathcal{A}^{\prime}_{\ell mn}),
    \end{align}
\end{subequations}
and the corresponding basis functions $v^I_{j,\ell mn}(t)$ are given by
\begin{subequations}
    \begin{align}
        v^I_{0,\ell mn}(t)&=F^I_+e^{-(t-t^I_{\mathrm{S}})/\tau_{\ell mn}}\cos[\omega_{\ell mn}(t-t^I_{\mathrm{S}})], \\
        v^I_{1,\ell mn}(t)&=F^I_+e^{-(t-t^I_{\mathrm{S}})/\tau_{\ell mn}}\sin[\omega_{\ell mn}(t-t^I_{\mathrm{S}})], \\
        v^I_{2,\ell mn}(t)&=F^I_{\times}e^{-(t-t^I_{\mathrm{S}})/\tau_{\ell mn}}\cos[\omega_{\ell mn}(t-t^I_{\mathrm{S}})], \\
        v^I_{3,\ell mn}(t)&=F^I_{\times}e^{-(t-t^I_{\mathrm{S}})/\tau_{\ell mn}}\sin[\omega_{\ell mn}(t-t^I_{\mathrm{S}})],
    \end{align}
\end{subequations}
where $t^I_{\mathrm{S}}$ represents the analysis start time at the $I$th detector.

We perform Bayesian inference based on the semianalytic method introduced in Ref.~\cite{Morisaki:2025gyu}, which uses orthonormalized basis vectors constructed from the waveform model described above.
Specifically, we orthonormalize the basis vectors $\{v^I_{j,\ell mn}\}$ with respect to the noise-weighted inner product using the Gram-Schmidt procedure, obtaining a new set of basis vectors $\{\tilde{v}^I_{j,\ell mn}\}$.
The strain is then expanded in this orthonormalized basis as
\begin{align}
    h^I(t)&=\sum_{\ell mn}\sum_{j=0}^3 \tilde{c}_{j,\ell mn}\tilde{v}^I_{j,\ell mn}(t),
\end{align}
where $\tilde{c}_{j,\ell mn}$ are the expansion coefficients.

In previous analyses~\cite{Isi:2021iql, Siegel:2024jqd, Isi:2019aib, Cotesta:2022pci, LIGOScientific:2025rid, LIGOScientific:2025wao, Siegel:2025xgb}, a uniform prior was imposed on the \qnm amplitudes
\begin{align}
    A_{\ell mn}&\equiv |\mathcal{A}_{\ell mn}|+|\mathcal{A}^{\prime}_{\ell mn}| \notag\\
    &=\frac{1}{2}\left[
        \sqrt{(c_{0,\ell mn}+c_{3,\ell mn})^2+(c_{1,\ell mn}-c_{2,\ell mn})^2}
    \right. \notag\\
    &\hspace{-0em}
    \left.
        +\sqrt{(c_{0,\ell mn}-c_{3,\ell mn})^2+(c_{1,\ell mn}+c_{2,\ell mn})^2}
    \right],
\end{align}
and the presence of a mode was inferred from whether the posterior distribution of $A_{\ell mn}$ excludes zero.
In contrast, in our method, a uniform prior is imposed on the orthonormalized \qnm amplitudes,
\begin{align}
    \tilde{A}_{\ell mn}\equiv\sqrt{\sum_{j=0}^3(\tilde{c}_{j,\ell mn})^2}, \label{eq:orthonormal_amplitude}
\end{align}
and the presence of a mode is inferred from whether the posterior distribution of $\tilde{A}_{\ell mn}$ excludes zero.
We, therefore, adopt $\tilde{A}_{\ell mn}$ as the indicator for the presence of a physical Kerr $(\ell,m,n)$ mode.
Consequently, while the prior is uniform in $\tilde{A}_{\ell mn}$, it generally induces a nonuniform effective prior on the physical \qnm amplitudes $A_{\ell mn}$.
To assess the presence of a mode using this indicator, we estimate the posterior density of $\tilde{A}_{\ell mn}$ using a reflected \kde and compute the smallest \hpd credible region that contains $\tilde{A}_{\ell mn}=0$, following the procedure adopted in Ref.~\cite{LIGOScientific:2025rid}.
The exclusion probability is defined as 1 minus the probability mass enclosed by this \hpd region.
Since the resulting probability depends on the \kde bandwidth, we adopt Scott's rule for bandwidth selection, consistent with Ref.~\cite{LIGOScientific:2025rid}.

We emphasize that $\tilde{A}_{\ell mn}$ is interpreted as the amplitude of the part of the Kerr $(\ell,m,n)$ mode orthogonal to all modes orthonormalized prior to the $(\ell,m,n)$ mode, rather than as the physical amplitude of the Kerr mode itself.
Our inference is then based on the property that
\begin{align}
    \tilde{A}_{\ell mn}=0~\Longleftrightarrow~A_{\ell mn}=0, \label{eq:mode_equivalence}
\end{align}
so that the presence (absence) of an orthonormal mode is equivalent to the presence (absence) of the corresponding Kerr \qnm.
Consequently, the degree to which the posterior distribution excludes $\tilde{A}_{\ell mn}=0$ can be used as an indicator for the existence of the corresponding Kerr \qnm.
This equivalence in Eq.~\eqref{eq:mode_equivalence} holds provided that the \qnms are orthonormalized in the correct order of significance.
In practice, this requires only that a mode absent from the data is not assigned a higher significance than a mode that is actually present.
A demonstration and discussion of the potential bias arising from an incorrect ordering is given in Appendix \ref{append:injection_tests}.

To test deviations from \gr, we allow the frequency and damping rate of the first overtone ($f_{221}\equiv\omega_{221}/2\pi$ and $\gamma_{221}\equiv1/\tau_{221}$) to differ from their Kerr values [$f^{\mathrm{(Kerr)}}_{221}(M_f,\chi_f)$ and $\gamma^{\mathrm{(Kerr)}}_{221}(M_f,\chi_f)$] in the 220+221 model.
We parametrize these deviations as $f_{221}=f^{\mathrm{(Kerr)}}_{221}(M_f,\chi_f)\exp(\delta f_{221})$ and $\gamma_{221}=\gamma^{\mathrm{(Kerr)}}_{221}(M_f,\chi_f)\exp(\delta \gamma_{221})$, where $\delta f_{221}=\delta\gamma_{221}=0$ corresponds to the Kerr prediction.
We then construct the basis vectors $\{v^I_{j,\ell mn}\}$ using these modified frequencies and damping rates, and obtain the orthonormalized basis $\{\tilde{v}^I_{j,\ell mn}\}$ following the same orthonormalization procedure described above.
Using this basis, we evaluate the likelihood of Eq.~(41) of Ref.~\cite{Morisaki:2025gyu}, which is analytically marginalized over the coefficients $\{\tilde{c}_{j,\ell mn}\}$.
We then perform Bayesian inference on the parameters $(M_f, \chi_f, \delta f_{221}, \delta \gamma_{221})$ based on this likelihood.

\begin{figure*}[htbp]
    \centering
    \begin{minipage}{0.495\textwidth}
        \includegraphics[width=\linewidth]{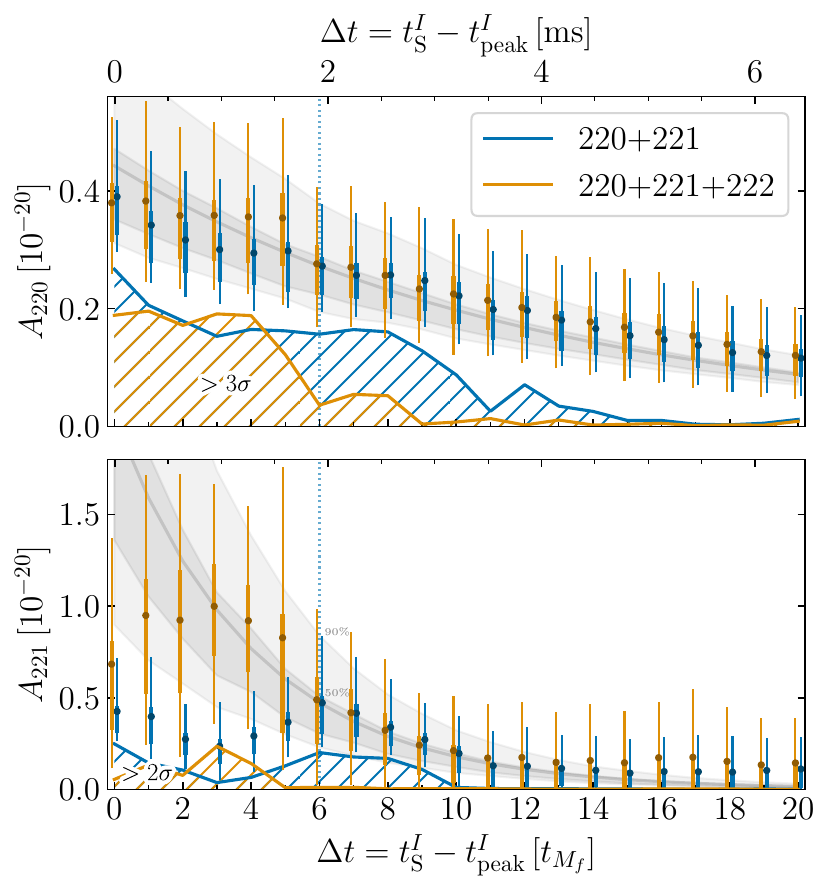}
    \end{minipage}
    \hfill
    \begin{minipage}{0.495\textwidth}
        \includegraphics[width=\linewidth]{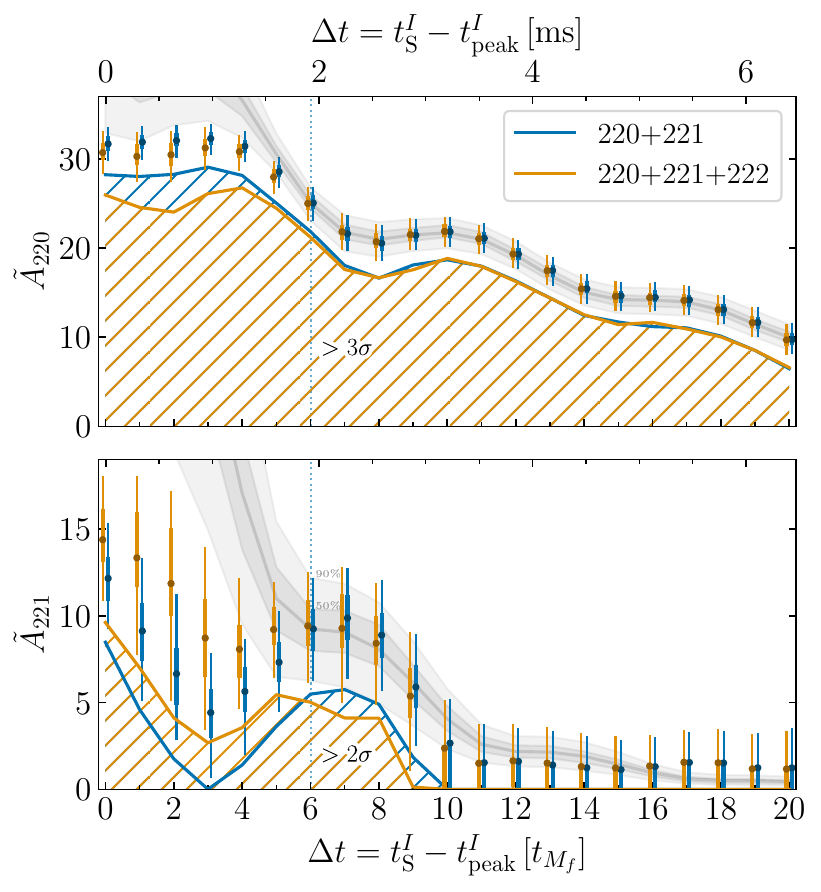}
    \end{minipage}
    \caption{
        Posterior distributions of the amplitudes of the 220 and 221 modes for the 220+221 (blue) and 220+221+222 (orange) models, evaluated for different analysis start times $\Delta t$.
        The left (right) panel shows the results from \textsc{ringdown} (the semianalytic method).
        The circles, thick vertical lines, and thin vertical lines represent the medians, 50\%, and 90\% credible intervals, respectively.
        The gray lines and shaded regions represent the means, 50\%, and 90\% credible intervals of the amplitudes predicted from the posterior of the 220+221 model at $\Delta t=6\,t_{M_f}$ (vertical dotted line), assuming exponential decay with the \qnm damping time.
        Note that, in the orthonormal analysis (right panel), the basis functions no longer correspond to simple exponentially damped sinusoids.
        As a result, the amplitudes constructed from their coefficients [see Eq.~\eqref{eq:orthonormal_amplitude}] exhibit oscillatory behavior.
        The hatched regions indicate exclusion at the $3\sigma$ (top panel) and $2\sigma$ (bottom panel) credible levels.
    }
    \label{fig:violin_plots}
\end{figure*}
\subsection{Setup}

For the complex frequencies $\tilde{\omega}_{\ell mn}$ used in the signal model for parameter inference, we adopt those of Kerr \qnms.
We compute them by interpolating the tabulated results of Refs.~\cite{Motohashi:2024fwt, motohashi_2025_14380191}, which provide precise calculations of the complex frequencies at discrete spin values.

We follow the choice of reference parameters in Ref.~\cite{LIGOScientific:2025rid}.
Specifically, we adopt the reference peak time $t_{\mathrm{peak}}=1420878141.235932\,\mathrm{s}$ (GPS) at geocenter, the reference remnant mass $M_f=68.409\,M_{\odot}$, the source right ascension $\alpha=2.333\,\mathrm{rad}$, declination $\delta=0.190\,\mathrm{rad}$, and polarization angle $\psi=0.190\,\mathrm{rad}$.
Given the reference peak time and sky location, the reference peak times at LIGO Hanford and Livingston detectors are $t^{\mathrm{H1}}_{\mathrm{peak}}=1420878141.2190118\,\mathrm{s}$ (GPS) and $t^{\mathrm{L1}}_{\mathrm{peak}}=1420878141.2165165\,\mathrm{s}$ (GPS), respectively.
We perform analyses using data segments starting at different times, characterized by $\Delta t\equiv t^I_{\mathrm{S}}-t^I_{\mathrm{peak}}$, expressed in units of $t_{M_f}\equiv GM_f/c^3=0.337\,\mathrm{ms}$.
In the analyses using the semianalytic method, we assume flat priors on the mass and spin, $M_f\in[40,100]\,M_{\odot}$ and $\chi_f\in[0,0.99]$.
For comparison, we perform Bayesian inference with the \textsc{ringdown} package~\cite{Isi:2021iql}, adopting the same priors on the mass and spin, and flat priors on the amplitudes $A_{\ell mn}\in[0,5\times10^{-20}]$.
In the analyses including deviations from \gr, we adopt flat priors on the deviation parameters, $\delta f_{221}\in[-0.8,0.8]$ and $\delta\gamma_{221}\in[-0.5,0.5]$.

The data conditioning follows the procedure adopted in Ref.~\cite{LIGOScientific:2025rid} and implemented in the \textsc{ringdown} package.
Specifically, we apply a 10\,Hz high-pass Butterworth filter to 634\,s of data around the peak time.
For each choice of the analysis start time $t^I_{\mathrm{S}}$, we identify the sample closest to $t^I_{\mathrm{S}}$ in the data sampled at the native LIGO sampling rate of 16384\,Hz.
The data are then downsampled to 4096\,Hz while preserving this sample.
We analyze 0.6\,s of the resulting data starting at $t^I_{\mathrm{S}}$.

The construction of the noise covariance matrix also follows the procedure adopted in Ref.~\cite{LIGOScientific:2025rid} and implemented in the \textsc{ringdown} package.
The \psd is taken from Ref.~\cite{LIGOScientific:2025rid} and modified by inflating its values below 20\,Hz and above 1830\,Hz to sufficiently large values.
The covariance matrix is then constructed in the frequency domain from this modified \psd and is used to define the noise-weighted inner product in the likelihood.

\begin{figure*}[htbp]
    \centering
    \begin{minipage}{0.495\textwidth}
        \includegraphics[width=\linewidth]{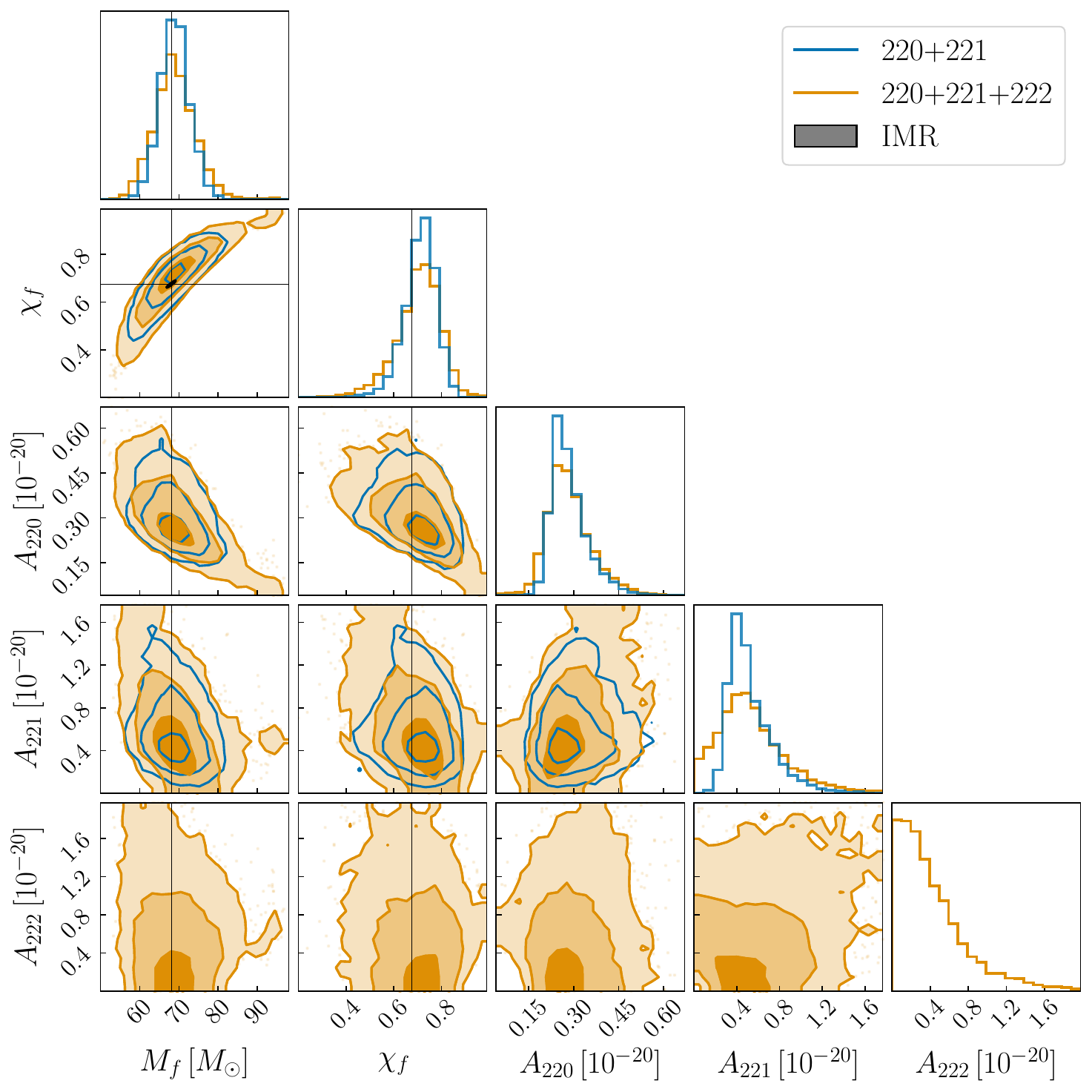}
    \end{minipage}
    \hfill
    \begin{minipage}{0.495\textwidth}
        \includegraphics[width=\linewidth]{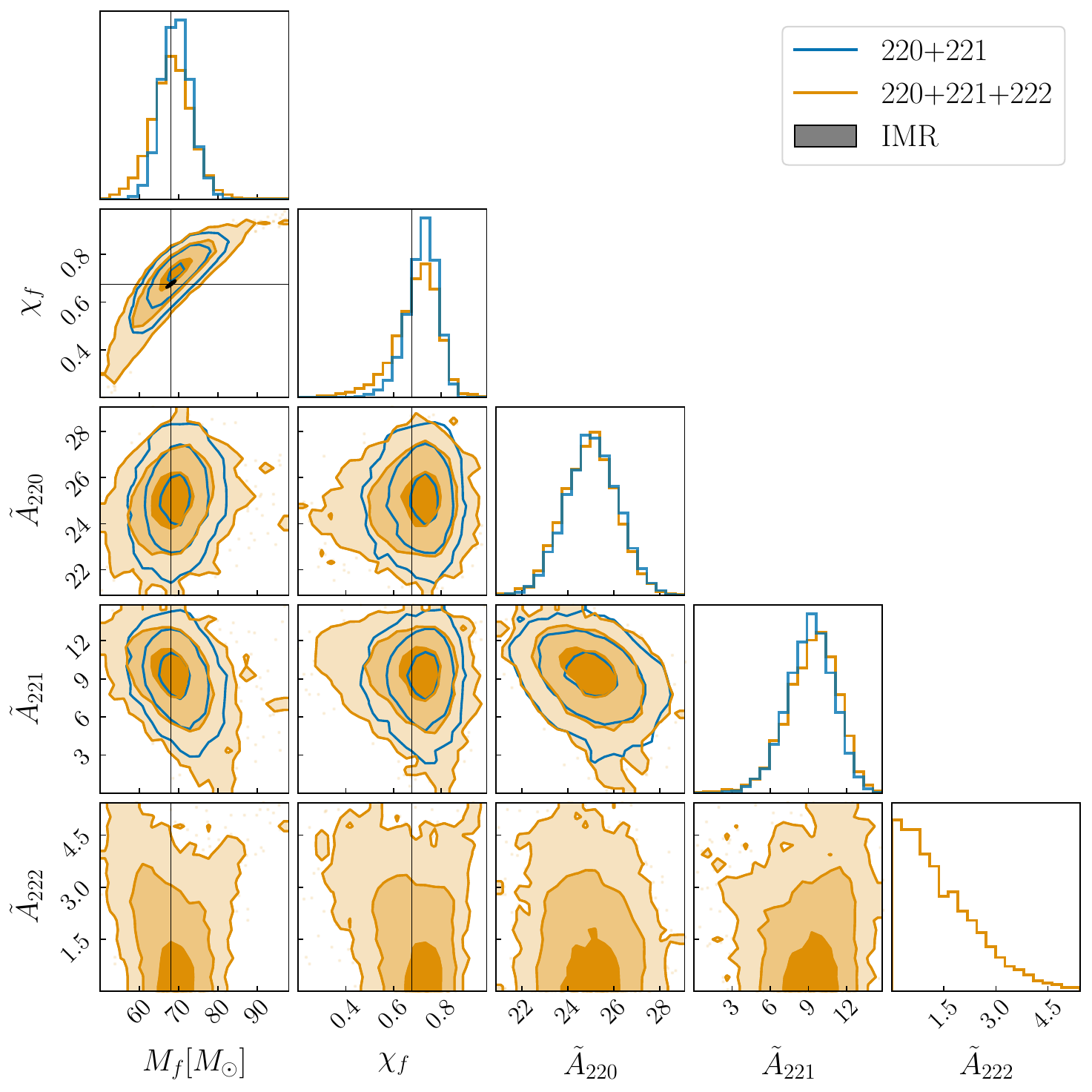}
    \end{minipage}
    \caption{
        Posterior distributions for the remnant \bh's mass, spin, and mode amplitudes for the 220+221 (blue) and 220+221+222 (orange) models, evaluated at $\Delta t=6\,t_{M_f}$.
        The left (right) panel shows the results from the nonorthogonal (orthonormal) analysis.
        Contours indicate the $1\sigma$ (39.3\%), $2\sigma$ (86.5\%), and $3\sigma$ (98.9\%) credible regions.
        Black contours show the 90\% credible regions obtained from the \imr analysis using \textsc{NRSur7dq4}~\cite{Varma:2019csw}, with black lines indicating their mean values ($M_f=68.1\,M_{\odot}$, $\chi_f=0.68$).
    }
    \label{fig:corner_plots}
\end{figure*}
\section{Results} \label{sec:results}

Using the above analysis setup, we analyze the GW250114 data~\cite{LIGOScientific:2025snk, KAGRA:2023pio, LIGOScientific:2019lzm} for various analysis start times $\Delta t$ in the range $\Delta t\in[0,20]\,t_{M_f}$.
For comparison with the semianalytic method, we perform the parameter estimation using the original (nonorthogonal) \qnms with the \textsc{ringdown} package~\cite{Isi:2021iql}.
We refer to this as the nonorthogonal analysis, while the semianalytic method based on orthonormalized \qnms is referred to as the orthonormal analysis.
In both analyses, we employ models including the $(\ell,m,n)=(2,2,0)$ and $(2,2,1)$ modes (220+221 model), as well as the $(2,2,0)$, $(2,2,1)$, and $(2,2,2)$ modes (220+221+222 model).

Figure~\ref{fig:violin_plots} shows the posterior distributions of the amplitudes of the $(2,2,0)$ and $(2,2,1)$ modes evaluated for different analysis start times $\Delta t$.
The blue distributions correspond to the results obtained with the 220+221 model, while the orange distributions correspond to those obtained with the 220+221+222 model.
In the nonorthogonal analyses (left panel), the amplitude of the $(2,2,1)$ mode in the 220+221 model is bounded away from zero for $\Delta t\leq9\,t_{M_f}$.
Furthermore, it exhibits an exponential decay consistent with the prediction using the result at $\Delta t=6\,t_{M_f}$ as a reference.
However, at earlier times than this reference time, it deviates from the prediction.
On the other hand, the 220+221+222 model slightly mitigates these deviations.
These results are broadly consistent with the results in Refs.~\cite{LIGOScientific:2025rid, LIGOScientific:2025wao}.

Including the $(2,2,2)$ mode extends the time range over which the amplitudes are consistent with the predicted exponential decay toward earlier times.
However, the resulting posterior distributions no longer clearly exclude zero, even when the 220+221 model yields amplitudes that are bounded away from zero ($\Delta t\in[5, 9]\,t_{M_f}$).
This behavior can be attributed to the correlation between the $(2,2,1)$ and $(2,2,2)$ mode amplitudes.
Figure~\ref{fig:corner_plots} shows the posterior distributions of the remnant \bh's mass, spin, and mode amplitudes at $\Delta t=6\,t_{M_f}$.
The blue distributions correspond to the results obtained with the 220+221 model, while the orange distributions correspond to those obtained with the 220+221+222 model.
From the two-dimensional posterior of the $(2,2,1)$ and $(2,2,2)$ mode amplitudes in the left panel, a negative correlation is observed.
As a result, even though the contribution of the $(2,2,2)$ mode is expected to be negligible, including it in the model alters the posterior distribution of the $(2,2,1)$ mode amplitude.
We explicitly demonstrate this correlation through an injection test in Appendix~\ref{append:injection_tests}.
While the 220+221 model clearly excludes a zero amplitude for the $(2,2,1)$ mode in the one-dimensional posterior, the exclusion of $A_{221}=0$ is reduced to 86.2\% in the 220+221+222 model.

On the other hand, the right panel of Fig.~\ref{fig:corner_plots}, which shows the results obtained using the orthonormal analysis, indicates that the correlation between the $(2,2,1)$ and $(2,2,2)$ mode amplitudes is reduced.
As a result, distributions of the $(2,2,0)$ and $(2,2,1)$ mode amplitudes are not significantly changed by including the $(2,2,2)$ mode in the model, and the one-dimensional posterior excludes $\tilde{A}_{221}=0$ at the 99.9\% level even in the 220+221+222 model.
This increased significance results from both the reduced correlations achieved by the reparametrization and the effective prior induced by adopting a uniform prior on $\tilde{A}_{\ell mn}$, which favors larger values of the physical amplitudes $A_{220}$ and $A_{221}$.
This effective prior is not introduced to artificially enhance the significance for the presence of the mode, rather, it arises naturally from our choice of $\tilde{A}_{\ell mn}$ as the indicator of the presence of a physical Kerr QNM mode.

As seen in Fig.~\ref{fig:violin_plots}, for $\Delta t\geq 6\,t_{M_f}$, the posterior distributions of the $(2,2,1)$ mode amplitude are not significantly affected by including the $(2,2,2)$ mode, compared to the results obtained with the nonorthogonal analysis.
In the top panel, the $(2,2,0)$ mode amplitude is consistent among the 220+221 model, the 220+221+222 model, and the prediction.
Moreover, the orthonormal analysis extends the time interval over which the (2,2,0) mode amplitude excludes zero at the 3$\sigma$ level compared to the nonorthogonal analysis.
At earlier times, however, the results of the two models show discrepancies and deviate from the prediction.
This may indicate the presence of additional contributions not included in the models, such as higher-order linear \qnms, non-modal signals~\cite{Oshita:2025qmn, Lu:2025vol}, and nonlinear \qnms~\cite{London:2014cma, Cheung:2022rbm, Mitman:2022qdl, Ma:2024qcv, Yang:2025ror, Wang:2026rev}.

\begin{figure*}[t]
    \centering
    \begin{minipage}{0.48\textwidth}
        \centering
        \includegraphics[width=\linewidth]{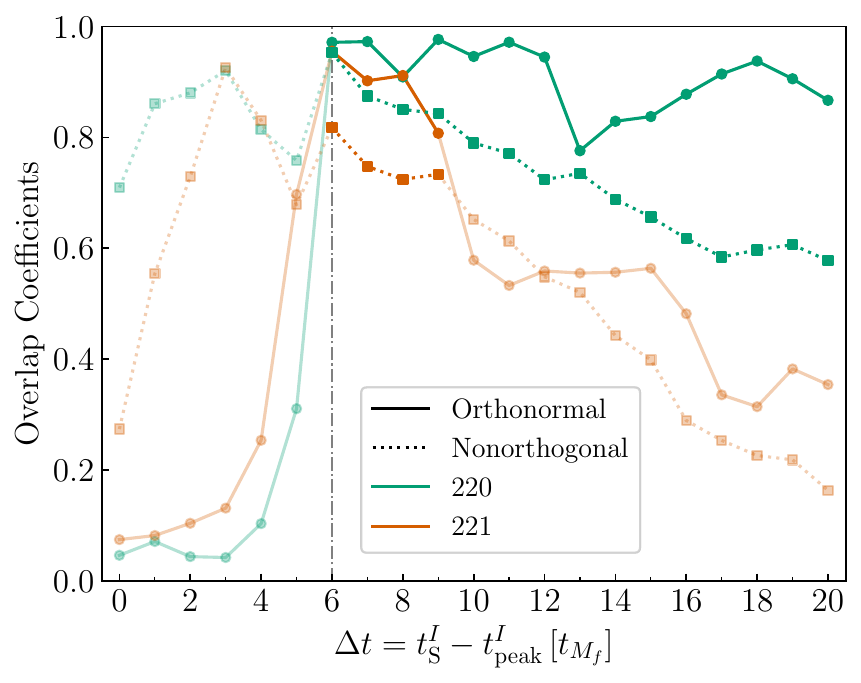}
        \caption{
            \ovls between the posterior distributions of \qnm amplitudes obtained from the 220+221+222 model and those predicted from the 220+221 fit at $\Delta t=6\,t_{M_f}$ (black dash-dotted line).
            Circles and squares denote the \ovls obtained from the orthonormal and nonorthogonal analyses, respectively.
            Green and red markers correspond to the $(2,2,0)$ and $(2,2,1)$ modes, respectively.
            The darker markers correspond to the time points after the reference time $6\,t_{M_f}$ at which the 220+221 model in the orthonormal analysis indicates the presence of the $(2,2,0)$ and $(2,2,1)$ modes at the $3\sigma$ and $2\sigma$ levels, respectively.
        }
        \label{fig:ovls}
    \end{minipage}
    \hfill
    \begin{minipage}{0.48\textwidth}
        \centering
        \includegraphics[width=\linewidth]{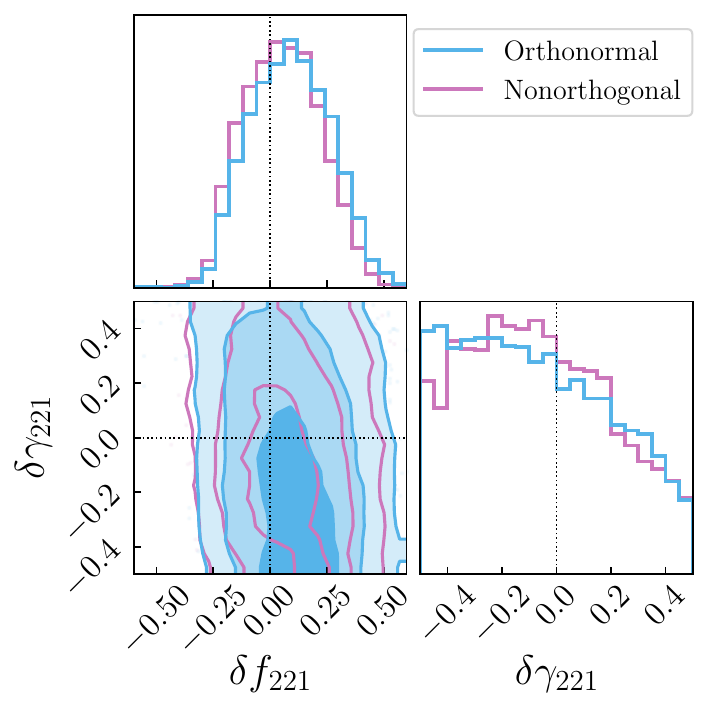}
        \caption{
            Posterior distributions of the deviation parameters $\delta f_{221}$ and $\delta\gamma_{221}$ obtained from the orthonormal analysis (light blue) and the nonorthogonal analysis (purple).
            The black dotted lines indicate the Kerr prediction, $\delta f_{221}=\delta\gamma_{221}=0$.
            The remnant mass $M_f$ and spin $\chi_f$ are marginalized over.
        }
        \label{fig:dfg}
    \end{minipage}
\end{figure*}
From the above results, we conclude that the orthonormal analysis provides a robust approach to detect \qnms, regardless of the number of modes included in the template.
To quantify the robustness of our method, we compute the \ovls between the distributions of \qnm amplitudes obtained from the 220+221+222 model and the prediction based on the 220+221 fit at $\Delta t=6\,t_{M_f}$.
Here, the \ovl between two probability distributions $p(x)$ and $q(x)$ is defined as
\begin{equation}
\mathrm{OVL} = \int\min\left[p(x),\,q(x)\right]\,dx,
\end{equation}
which takes values between 0 and 1, with values closer to unity indicating a higher degree of agreement between the two distributions.
Figure~\ref{fig:ovls} shows the \ovls obtained from the orthonormal analysis (solid lines) and the nonorthogonal analysis (dotted lines).
Here, larger \ovl values indicate better agreement between the posterior distributions obtained from the 220+221+222 model and the decay behavior predicted from the 220+221 model.
The darker markers correspond to the time points after the reference time at which the orthonormal analysis with the 220+221 model indicates the presence of the $(2,2,0)$ and $(2,2,1)$ modes at the $3\sigma$ and $2\sigma$ levels, i.e., $6\,t_{M_f}\leq\Delta t$ and $6\,t_{M_f}\leq\Delta t\leq9\,t_{M_f}$, respectively.
For both the (2,2,0) and (2,2,1) modes, the orthonormal analysis yields larger \ovls than the nonorthogonal analysis at the time points indicated by the darker markers. 
Specifically, the mean \ovl over these time points is 95.8\% for the orthonormal analysis and 88.0\% for the nonorthogonal analysis for the (2,2,0) mode, and 89.4\% and 75.6\%, respectively, for the (2,2,1) mode.

Finally, we introduce the deviation parameters $\delta f_{221}$ and $\delta\gamma_{221}$ in the 220+221 model.
We perform the analysis including these parameters fixing the analysis start time to $\Delta t=6\,t_{M_f}$.
As in the previous analyses, we carry out the Bayesian inference using both the orthonormal analysis and the nonorthogonal analysis using the \textsc{ringdown} package for comparison.

Figure~\ref{fig:dfg} shows the posterior distributions of the deviation parameters $\delta f_{221}$ and $\delta \gamma_{221}$.
The light blue distributions correspond to the results obtained from the orthonormal analysis, while the purple distributions correspond to those obtained from the nonorthogonal analysis.
As reported in Ref.~\cite{LIGOScientific:2025rid}, no significant constraint is obtained on $\delta\gamma_{221}$.
The orthonormal analysis yields $\delta f_{221}=0.09^{+0.29}_{-0.27}$, while the nonorthogonal analysis gives $\delta f_{221}=0.05^{+0.27}_{-0.26}$ (90\% credible intervals), both of which are consistent with Ref.~\cite{LIGOScientific:2025rid}.
In both analyses, no significant deviation from the Kerr prediction is observed.

\section{Conclusion}

In this paper, we perform a ringdown analysis of GW250114 using a semianalytic method based on orthonormalized \qnms (orthonormal analysis), and compare the results with those obtained from a conventional analysis based on nonorthogonal \qnms with the \textsc{ringdown} package~\cite{Isi:2021iql} (nonorthogonal analysis).
We consider models including the $(\ell,m,n)=(2,2,0)$ and $(2,2,1)$ modes (220+221 model), as well as an extended model incorporating the $(2,2,2)$ mode (220+221+222 model).
Due to the high \snr of the GW250114 ringdown, the posterior distributions of the amplitudes of not only the dominant $(2,2,0)$ mode but also the subdominant $(2,2,1)$ mode remain bounded away from zero up to $9\,t_{M_f}$ after the merger in the 220+221 model.
However, when the $(2,2,2)$ mode is included, the amplitude of the $(2,2,1)$ mode is no longer bounded away from zero in the nonorthogonal analysis, reflecting the correlation between the $(2,2,1)$ and $(2,2,2)$ modes.
On the other hand, the orthonormal analysis mitigates this correlation and leaves the amplitudes largely unchanged across both models.
As a result, in the 220+221+222 model, the amplitude of the $(2,2,1)$ mode excludes zero at the 99.9\% level at $6\,t_{M_f}$ after the merger, compared to 86.2\% in the nonorthogonal analysis.
We also perform a test of \gr by allowing for deviations in the frequency and damping rate of the $(2,2,1)$ mode.
Both the orthonormal and nonorthogonal analyses yield consistent results and no significant deviation from the Kerr prediction is observed.

The above results on the mode amplitudes demonstrate that the orthonormal analysis enables more robust detection of the subdominant $(2,2,1)$ mode, regardless of the number of \qnms included in the template waveform.
This robustness is expected to be especially important in cases where the correlations between modes are stronger.
A future high-\snr event in which multiple subdominant modes are detectable is a natural example.
Such events are expected to become accessible in the era of the upcoming fifth \lvk observing run and next-generation detectors such as the Cosmic Explorer~\cite{Reitze:2019iox} and the Einstein Telescope~\cite{Punturo:2010zz}, where our method will become more powerful.

Another interesting example is a high-spin Kerr ringdown, where the \qnm frequencies of certain overtones become closely spaced, and these overtones are resonantly excited~\cite{Motohashi:2024fwt, Yang:2025dbn}.
In this regime, correlations among the overtone amplitudes are expected to become strong, making the orthonormal analysis potentially advantageous.
Exploring the effectiveness of our method in such cases is left for future work.

\acknowledgements

We thank Emanuele Berti and Alex B. Nielsen for useful discussions, Masaki Iwaya for reviewing this manuscript as part of the LIGO Publications and Presentations review process, and Juan Calderon Bustillo for helpful comments during the review process.
This work was supported by JST SPRING, Grant Number JPMJSP2108 (M.S.), and JSPS KAKENHI Grant Numbers JP23H04891 (S.M.), JP23H04893 (S.M.), JP22K03639 (H.M.), JP26K21813 (H.M.), and JP23KJ06945 (D.W.), and the Royal Society Award ICA\textbackslash R1\textbackslash 231114 (S. M).
The authors are grateful for computational resources provided by the LIGO Laboratory and supported by National Science Foundation Grants PHY-0757058 and PHY-0823459.

This research has made use of data or software obtained from the Gravitational Wave Open Science Center (gwosc.org), a service of the LIGO Scientific Collaboration, the Virgo Collaboration, and KAGRA. This material is based upon work supported by NSF's LIGO Laboratory which is a major facility fully funded by the National Science Foundation, as well as the Science and Technology Facilities Council (STFC) of the United Kingdom, the Max-Planck-Society (MPS), and the State of Niedersachsen/Germany for support of the construction of Advanced LIGO and construction and operation of the GEO600 detector. Additional support for Advanced LIGO was provided by the Australian Research Council. Virgo is funded, through the European Gravitational Observatory (EGO), by the French Centre National de Recherche Scientifique (CNRS), the Italian Istituto Nazionale di Fisica Nucleare (INFN) and the Dutch Nikhef, with contributions by institutions from Belgium, Germany, Greece, Hungary, Ireland, Japan, Monaco, Poland, Portugal, Spain. KAGRA is supported by Ministry of Education, Culture, Sports, Science and Technology (MEXT), Japan Society for the Promotion of Science (JSPS) in Japan; National Research Foundation (NRF) and Ministry of Science and ICT (MSIT) in Korea; Academia Sinica (AS) and National Science and Technology Council (NSTC) in Taiwan.

\section*{Data Availability}

The data that support the findings of this article are openly available~\cite{suzuki_2026_19939742}.

\begin{figure*}[htbp]
    \centering
    \begin{minipage}{0.495\textwidth}
        \includegraphics[width=\linewidth]{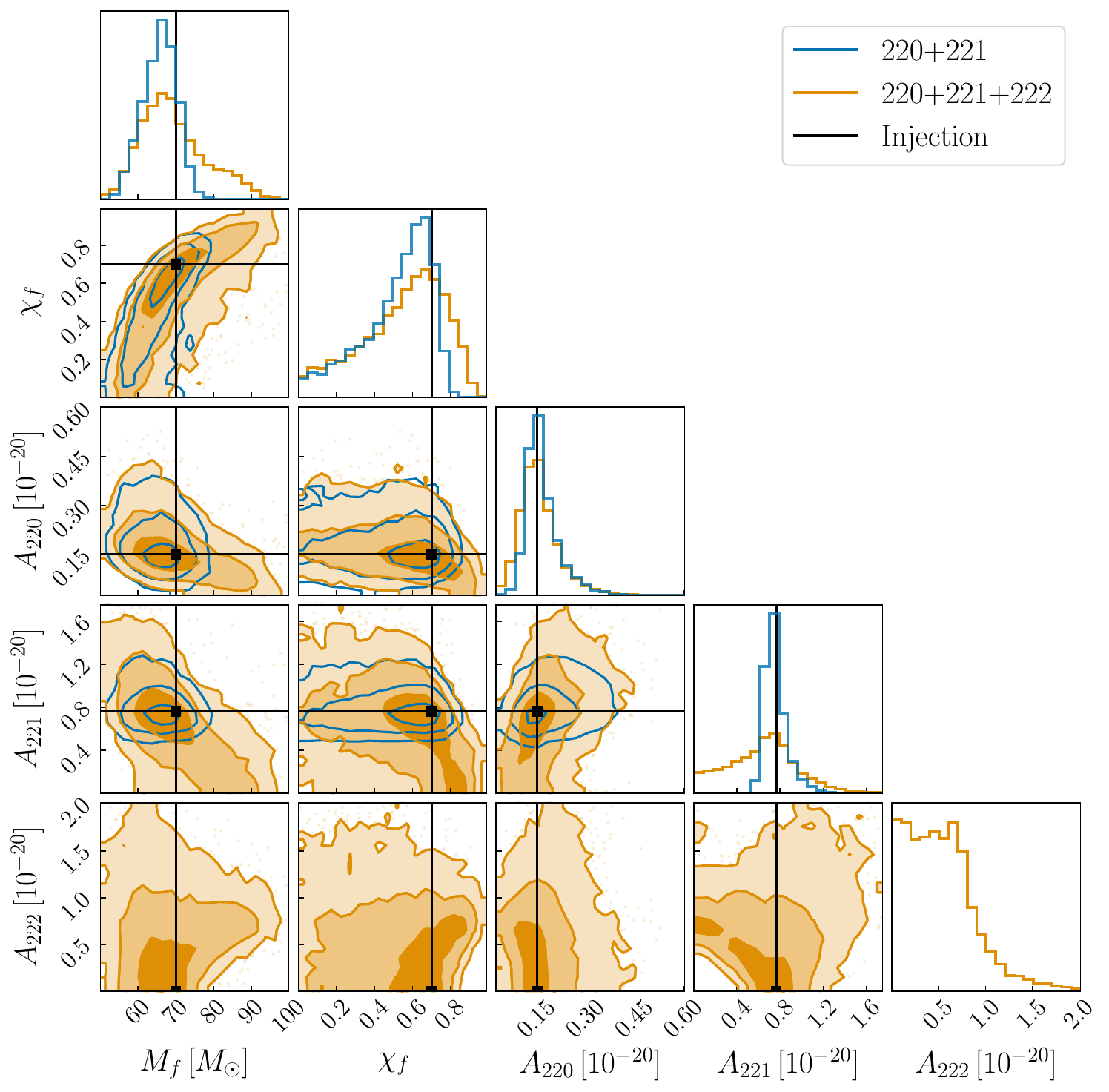}
    \end{minipage}
    \hfill
    \begin{minipage}{0.495\textwidth}
        \includegraphics[width=\linewidth]{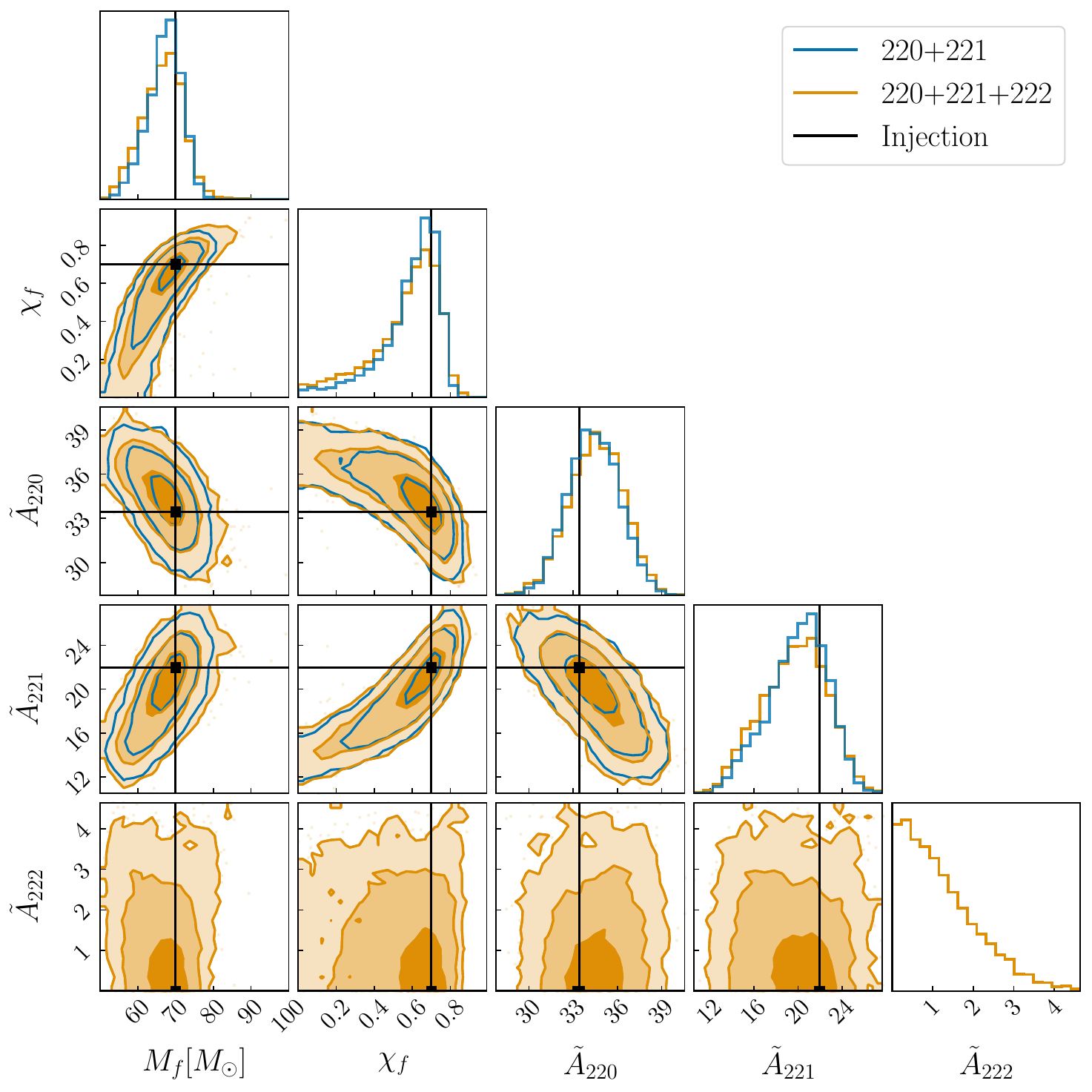}
    \end{minipage}
    \caption{
        Posterior distributions for the remnant \bh's mass, spin, and mode amplitudes for the 220+221 signal.
        Blue (orange) distributions correspond to the 220+221 (220+221+222) model.
        The left (right) panel shows the results from the nonorthogonal (orthonormal) analysis.
        Contours indicate the $1\sigma$ (39.3\%), $2\sigma$ (86.5\%), and $3\sigma$ (98.9\%) credible regions.
        Black lines indicate the injected values.
    }
    \label{fig:corner_plots_220+221}
\end{figure*}
\appendix
\section{Injection tests for mode identification} \label{append:injection_tests}
To assess the robustness of our method for mode identification and to investigate potential biases arising from the assumed mode hierarchy, we perform a set of injection tests.

\subsection*{Signal model and analysis setup}
We employ injection signals constructed from superposition of damped sinusoids given in Eq.~\eqref{eq:complex_strain}.
Specifically, we consider two cases: a signal composed of the (2,2,0) and (2,2,1) modes (220+221 signal), and a signal composed of the (2,2,0) and (2,2,2) modes (220+222 signal).
In both cases, the remnant mass and spin are set to $M_f=70\,M_{\odot}$ and $\chi_f=0.7$.
The complex amplitudes $\mathcal{A}_{\ell mn}$ and $\mathcal{A}^{\prime}_{\ell mn}$ are taken to be proportional to $B_{\ell mn}/(\tilde{\omega}_{\ell mn})^2$ and $(B_{\ell mn}/(\tilde{\omega}_{\ell mn})^2)^{\ast}$, following a simplified factorization motivated by \bh perturbation theory.
Here, $B_{\ell mn}$ is the excitation factor determined solely by $M_f$ and $\chi_f$, and we adopt the values computed in Ref.~\cite{motohashi_2025_14380191}.
The overall waveform is then rescaled such that the \snr measured from the peak is 40.
We thus employ the signal model
\begin{align}
    h(t)=N\sum_{n}\qty[
        \frac{B_{22n}}{\tilde{\omega}^2_{22n}}e^{-i\tilde{\omega}_{22n}t}
        +\qty(\frac{B_{22n}}{\tilde{\omega}^2_{22n}})^{\ast}e^{i\tilde{\omega}^{\ast}_{22n}t}
    ],
\end{align}
where $N$ is an overall normalization factor, chosen such that the \snr is 40.

For both the injection signals and the analysis model, we fix the peak time at the geocenter $t_{\mathrm{peak}}$, as well as the source right ascension $\alpha$, declination $\delta$, and polarization angle $\psi$, to the values used in the main analysis.
We also adopt the same priors as in the main analysis.

For the analysis data, we set the sampling rate to 4096\,Hz and the duration to 0.1\,s.
The noise covariance matrix used in the likelihood is identical to that used in the main analysis.
No random detector noise is added, and no filtering is applied.

\subsection*{Results}
Using the above analysis setup, we analyze the injection signals with the start time set at the signal peak.
As in the main analysis, we perform both nonorthogonal and orthonormal analyses.

Figure~\ref{fig:corner_plots_220+221} shows the posterior distributions of the remnant \bh's mass, spin, and mode amplitudes for the 220+221 signal.
The blue distributions correspond to the results obtained with the 220+221 model, while the orange distributions correspond to those obtained with the 220+221+222 model.
Consistent with the main analysis, a negative correlation is observed between the (2,2,1) and (2,2,2) mode amplitudes in the nonorthogonal analysis (left panel) for the 220+221+222 model.
As a result, the detection significance of the (2,2,1) mode is reduced when the (2,2,2) mode is included, even though the injected signal contains no (2,2,2) contribution.
On the other hand, in the orthonormal analysis (right panel), this correlation is reduced, and the (2,2,1) mode is detected with high significance not only in the 220+221 model but also in the 220+221+222 model.

\begin{figure*}[htbp]
    \centering
    \begin{minipage}{0.495\textwidth}
        \includegraphics[width=\linewidth]{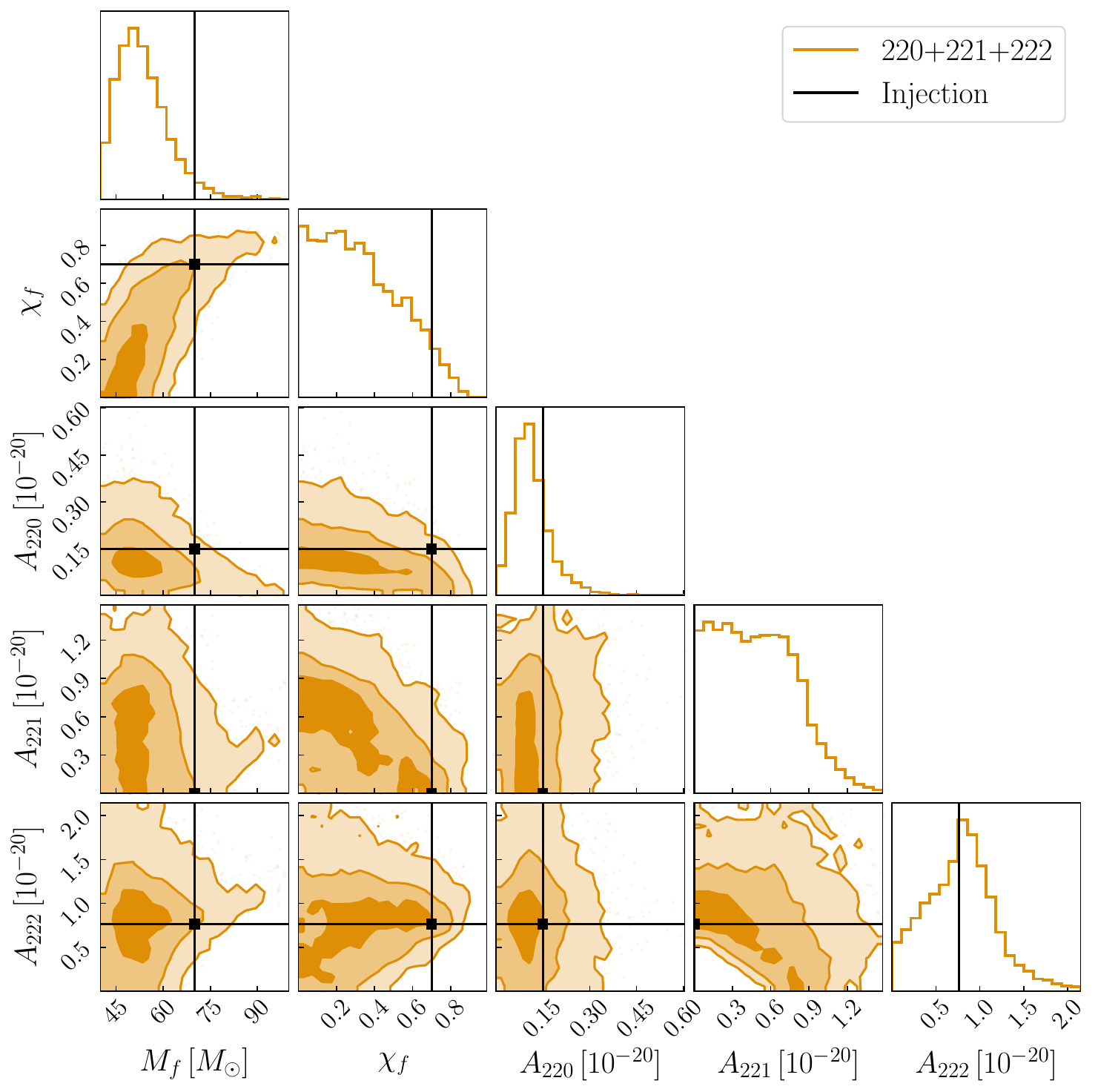}
    \end{minipage}
    \hfill
    \begin{minipage}{0.495\textwidth}
        \includegraphics[width=\linewidth]{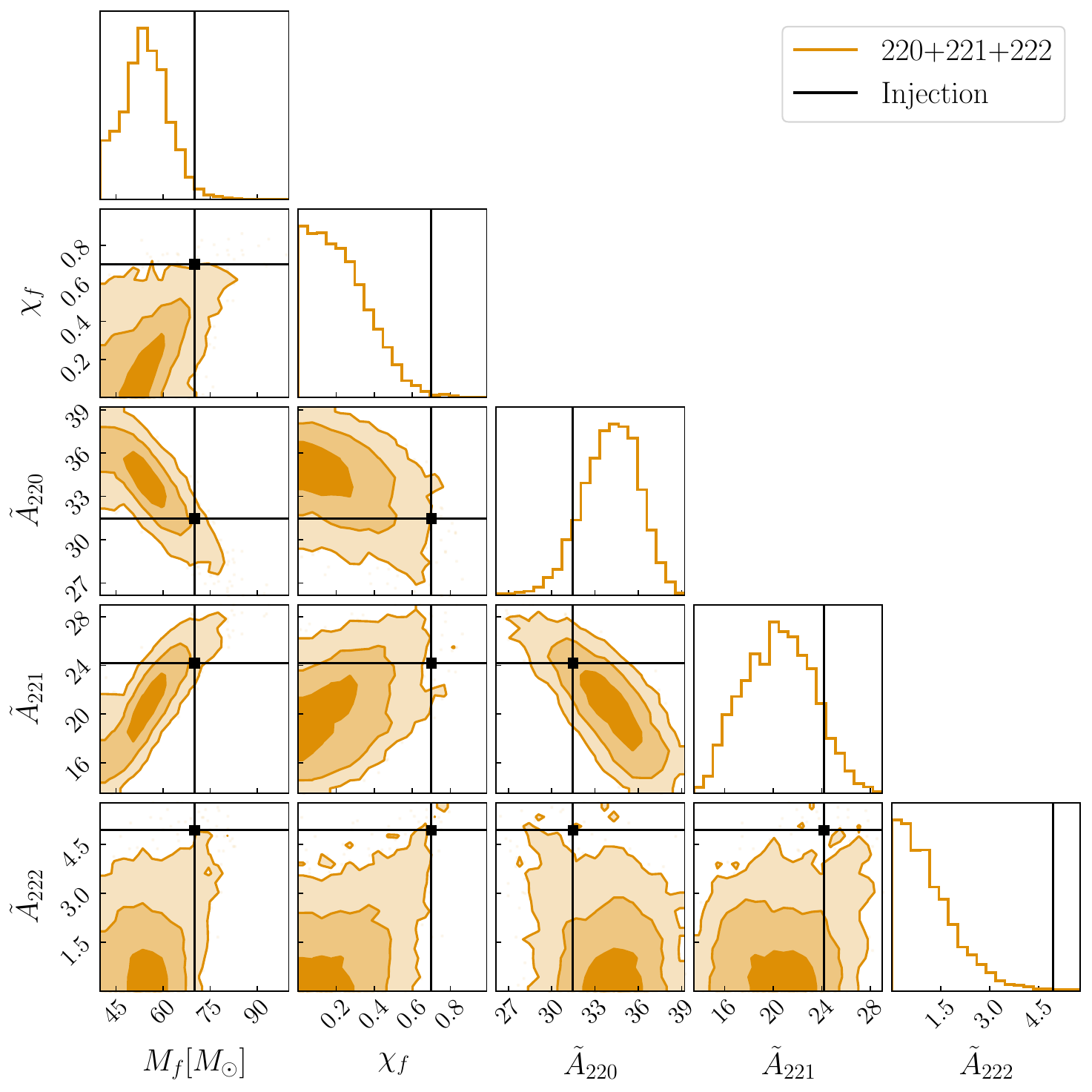}
    \end{minipage}
    \caption{
        Posterior distributions for the remnant \bh's mass, spin, and mode amplitudes for the 220+222 signal with the 220+221+222 model.
        The left (right) panel shows the results from the nonorthogonal (orthonormal) analysis.
        Contours indicate the $1\sigma$ (39.3\%), $2\sigma$ (86.5\%), and $3\sigma$ (98.9\%) credible regions.
        Black lines indicate the injected values.
    }
    \label{fig:corner_plots_220+222}
\end{figure*}
Next, we consider the more extreme case of the 220+222 signal, in which the assumed hierarchy of overtones is deliberately violated.
Figure~\ref{fig:corner_plots_220+222} shows the posterior distributions of the remnant \bh's mass, spin, and mode amplitudes for the 220+222 signal analyzed with the 220+221+222 model.
In this case, the nonorthogonal analysis (left panel) captures the true value of the (2,2,2) mode amplitude, while in the orthonormal analysis (right panel), the distributions of the (2,2,1) and (2,2,2) mode amplitudes are shifted away from their true values.

The shift in the posterior peak observed in the orthonormal analysis should not be interpreted as an intrinsic limitation of the method, but rather as a consequence of an incorrect modeling assumption.
The semianalytic method of Ref.~\cite{Morisaki:2025gyu} assumes that \qnms are ordered by their significance, and the orthonormalization is performed accordingly.
For instance, when the modes are ordered as (2,2,0), (2,2,1), and (2,2,2), the model effectively assumes the presence of either no modes, only the (2,2,0) mode, (2,2,0) and (2,2,1) modes, or all three modes
\footnote{
    That is, configurations with $(2,2,1)+(2,2,2)$ or $(2,2,0)+(2,2,2)$ are not considered.
}.
Under this assumption, there exists a one-to-one correspondence between the amplitudes of the nonorthogonal modes $A_{\ell mn}$ and those of the orthonormal modes $\tilde{A}_{\ell mn}$ in terms of whether they are finite or vanishing.
However, when this assumption is violated---for example, when the 220+222 signal is analyzed with the 220+221+222 model---this correspondence breaks down, as illustrated in the right panel of Fig.~\ref{fig:corner_plots_220+222}.
In such case, the orthonormal mode amplitudes no longer faithfully reflect the presence or absence of the corresponding nonorthogonal modes.
This can lead to situations where the (2,2,2) mode appears to be suppressed, even though it is present in the signal.

We therefore interpret this behavior as arising from a mismatch between the assumed mode hierarchy and the true signal content, rather than as a fundamental limitation of the orthonormal analysis itself.
In practice, the ordering of modes by significance can be reasonably informed by insights from the \imr analysis and fitting for numerical relativity.
For example, which $(\ell,m)$ modes are preferentially excited can be inferred from the \imr analysis~\cite{Zhu:2023fnf, Li:2021wgz}, while for modes with the same $(\ell,m)$ indices, their relative excitation is expected to follow a hierarchy largely determined by their excitation factors~\cite{Cheung:2023vki,Giesler:2024hcr,Mitman:2025hgy}.
Moreover, ordering modes by increasing the damping rate, as adopted in this work, is expected to provide a good approximation at sufficiently late times after the peak.
Such assumptions about mode hierarchy have been widely employed in previous analyses~\cite{Isi:2021iql, Isi:2019aib, LIGOScientific:2025rid, LIGOScientific:2025wao, Takahashi:2023tkb, Mitman:2025hgy, Clarke:2024lwi}.

\bibliography{references}

\end{document}